\documentclass[11pt]{iopart}
\pdfoutput=1
\usepackage{epsfig,graphicx,latexsym,iopams}

\newcommand\beq{\begin{equation}}
\newcommand\eeq{\end{equation}}
\newcommand{\ba}{\begin{eqnarray}}
\newcommand{\ea}{\end{eqnarray}}  
\def\spose#1{\hbox to 0pt{#1\hss}}
\def\lta{\mathrel{\spose{\lower 3pt\hbox{$\mathchar"218$}}
     \raise 2.0pt\hbox{$\mathchar"13C$}}}
\def\gta{\mathrel{\spose{\lower 3pt\hbox{$\mathchar"218$}}
     \raise 2.0pt\hbox{$\mathchar"13E$}}}

\def\msun{M_{\odot}}

\begin{document}

\title{Probing seed black holes using future gravitational-wave detectors}

\author{Jonathan R Gair}
\address{Institute of Astronomy, University of Cambridge, Cambridge, CB3 0HA, UK}
\ead{jgair@ast.cam.ac.uk}
\author{Ilya Mandel}
\address{Department of Physics and Astronomy, Northwestern University, Evanston, IL, 60208 USA}
\author{Alberto Sesana}
\address{Center for Gravitational Wave Physics, The Pennsylvania State University, University Park, PA16802, USA}
\author{Alberto Vecchio}
\address{School of Physics and Astronomy, University of Birmingham, Edgbaston, Birmingham, B15 2TT, UK}

\begin{abstract}
Identifying the properties of the first generation of seeds of massive black holes is key to understanding the merger history and growth of galaxies. Mergers between $\sim100M_{\odot}$ seed black holes generate gravitational waves in the $0.1$--$10$Hz band that lies between the sensitivity bands of existing ground-based detectors and the planned space-based gravitational wave detector, the Laser Interferometer Space Antenna (LISA). However, there are proposals for more advanced detectors that will bridge this gap, including the third generation ground-based Einstein Telescope and the space-based detector DECIGO. In this paper we demonstrate that such future detectors should be able to detect gravitational waves produced by the coalescence of the first generation of light seed black-hole binaries and provide information on the evolution of structure in that era. These observations will be complementary to those that LISA will make of subsequent mergers between more massive black holes. We compute the sensitivity of various future detectors to seed black-hole mergers, and use this to explore the number and properties of the events that each detector might see in three years of observation. For this calculation, we make use of galaxy merger trees and two different seed black hole mass distributions in order to construct the astrophysical population of events. We also consider the accuracy with which networks of future ground-based detectors will be able to measure the parameters of seed black hole mergers, in particular the luminosity distance to the source. We show that distance precisions of $\sim30\%$ are achievable, which should be sufficient for us to say with confidence that the sources are at high redshift.
\end{abstract}

\section{Introduction}
It is generally accepted that the massive black-holes (MBHs) found in the
centres of most galaxies grow via accretion and via mergers following
mergers between their host galaxies. However, little is known about the
seeds from which MBHs grow~\cite{SVH07}. The seeds could be {\it massive} (i.e., $\sim 10^5\msun$), in which case the first epoch of mergers between galactic black holes at high
redshift will generate gravitational waves (GWs) that could be detected by
the future space-borne detector LISA~\cite{lisa}. However, these seeds could also
be {\it light} (i.e., $\sim 100\msun$) and, in that case, the GWs generated during these first
mergers will be at frequencies between the sensitive bands of LISA and the
ground-based instruments currently in operation --- LIGO, Virgo and GEO 600~\cite{ligo,virgo,geo}, or their upgraded Advanced versions. In this paper we show that third-generation ground-based interferometers and second-generation space-based detectors may be able to fill this gap and directly probe the first mergers between light MBH seeds.

Light seed black holes might be produced as the remnants of pop III stars~\cite{SVH07,madau01}. Models including light seeds predict dozens of MBH binary (MBHB)
coalescences per year in the mass range $\sim10^2-10^6\msun$~\cite{sesana04}. Here and elsewhere we will stretch the definition of MBH to mean any black hole in the centre of a
dark matter halo. These events are mostly at high redshift, $z \gtrsim 3$, and
therefore only events with total mass $\gtrsim 10^3\msun$ will be accessible to
LISA. To observe systems in the $10-10^3\msun$ range will require a GW
detector sensitive to frequencies in the $0.1-10$Hz range. This might be
achieved in space by second-generation instruments such as the Big-Bang-Observer~\cite{bbo}, DECIGO~\cite{decigo} or ALIA~\cite{alia}, or on the ground by third-generation interferometers. In this paper we will use DECIGO as an example of a future space-based detector, and the Einstein gravitational-wave Telescope (ET)~\cite{hild08,freise09} as an example of a future ground-based detector. This choice was motivated by the fact that ET is currently undergoing a design study within Europe, and a technology demonstration mission for DECIGO is currently under development in Japan. DECIGO is a very ambitious mission requiring new technology and so the timescale is rather uncertain, but at present the launch target is $\sim2024$. The timetable for ET is also rather uncertain, but construction will start no earlier than $\sim 2018$. We will demonstrate that these instruments will have the capability to detect seed black-hole binaries, and we will discuss whether these observations could tell us definitively that we are observing light seeds produced by pop III stars. We have already estimated the seed black-hole event rates for ET~\cite{apjpaper} as input for the design study, but in this paper we compare and contrast these results with those for DECIGO, and provide additional details regarding the distribution of the detectable sources and parameter-estimation accuracy.

This paper is organised as follows. In Section~\ref{methods} we describe our calculations, including the astrophysical model used to construct the event populations, the waveform models used and the sensitivity curves of the various instruments. In Section~\ref{rates} we provide details of  the detectable seed black-hole merger events, including the number of events and the mass and redshift distributions of events. In Section~\ref{parest} we discuss the parameter estimation accuracy that might be achieved for these systems by a network of third-generation ground-based detectors, before summarizing our conclusions in Section~\ref{discuss}.

\section{Event rate calculation}
\label{methods}
\subsection{Astrophysical models for Pop III seed growth}
To generate event populations, we trace the galaxy
merger hierarchy using Monte-Carlo merger-tree realisations built on the
extended Press-Schechter formalism~\cite{press74}, using simulations described in detail
in~\cite{vhm,vsh}. In these simulations, dark matter halos are populated with $\sim 100\msun$
seed black holes at redshift $z \approx 20$, and it is assumed that these black holes
merge as their host halos merge, and accrete mass efficiently. In the
present work, we assumed a standard LCDM cosmology with WMAP 1-year
parameters~\cite{wmap} and considered two variants of the Pop III seed model. In both cases, we used the same merger history for the dark-matter halos and assumed that the seeds accrete at the Eddington limit during each merger episode, but the models differed in the initial mass-distribution of the seeds: (i) the {\it VHM,ems} (Volonteri-Haardt-Madau (VHM) with equal-mass seeds) model assumes equal $150\msun$ seeds; (ii) the {\it VHM,smd} model (VHM with seed-mass distribution) takes the seed-mass distribution to be uniform in log-mass for masses in the range $10-600\msun$. We have also studied two further scenarios that differ in the model of accretion onto the seeds, using prescriptions in~\cite{shank,hopk}. Results from all four scenarios were considered for the ET study in~\cite{apjpaper}, but we now restrict to the two representative models {\it VHM,ems} and {\it VHM,smd} to avoid overloading the figures in this paper.

These models produce event populations that are consistent with observational constraints at  $z<3$ from the X-ray and optical quasar luminosity function and from the observed faint X-ray counts of AGNs~\cite{vsh}. The models predict $\sim50$ MBHB coalesences per year in
the Universe when summed over all black hole masses, but uncertainties in the
assumptions of the model could change this by a factor of a few either way.
We are using these models to study black holes at $z>5$, but they have been tuned
to reproduce observations at lower redshift, $z < 3$. Indeed, accretion onto
light black holes may be very inefficient~\cite{alvarez08,milos08}, which adds further
uncertainties to the picture. However, we are using these specific models
only to indicate the potential of future GW observations as a probe of seed
black holes. Our results are not designed to be robust predictions of the
event rates given our present limited understanding of the details of the physical
processes involved.

\subsection{Signal-to-noise ratio and parameter-estimation accuracy calculations}
We compute signal-to-noise ratios (SNRs), $\rho$, for these sources using the usual expression, $\rho^2 = \langle h | h\rangle$, where $\langle \cdot | \cdot \rangle$ is the noise-weighted inner-product
\begin{equation}
\left<a\left|b\right.\right> =4 {\rm Re} \int_{0}^{\infty}\frac{\tilde{a}^*(f)\tilde{b}(f)}{S_{n}(f)}\,{\rm d}f
\label{snr}
\end{equation}
in which $\tilde{h}(f)$ is the Fourier transform of the waveform, and $S_n(f)$ is the one-sided power spectral density of the noise in the detector. To assess parameter-estimation accuracies, we compute the inverse of the Fisher information matrix (FIM)
\begin{equation}
\Gamma_{ij} = \langle \frac{\partial h}{\partial \lambda_i} |  \frac{\partial h}{\partial \lambda_j} \rangle,
\end{equation}
where $\lambda_i$ denotes the model parameters. When considering multiple detectors in a network, the above expressions are still valid if the inner product is replaced by the sum of the inner products over the individual detectors.

\subsection{Waveform  model}
To model the gravitational waves generated by MBHB mergers, we use the phenomenological waveform model (IMR) described in~\cite{ajith}. This waveform model includes the inspiral, merger and ringdown radiation in a consistent way using a prescription of the form
\begin{equation}
\fl
u(f) \equiv A_{\rm eff}(f) \exp\left({\rm i}\Psi_{\rm eff}(f)\right), \qquad A_{\rm eff} \equiv C \left\{ \begin{array}{ll} (f/f_{\rm merg})^{-7/6}&\mbox{if }f<f_{\rm merg} \\ (f/f_{\rm merg})^{-2/3}&\mbox{if }f_{\rm merg} \leq f<f_{\rm ring} \\ w{\cal L}(f,f_{\rm ring},\sigma)&\mbox{if }f_{\rm ring} \leq f<f_{\rm cut}\end{array}\right.
\end{equation}
Expressions for $f_{\rm merg}$, $f_{\rm ring}$, $f_{\rm cut}$, $\Psi_{\rm eff}(f)$, $C$, $w$ and ${\cal L}(f,f_{\rm ring},\sigma)$ are given in Eqs.~(4.14)-(4.19) and Tables I-II of~\cite{ajith}. For these sources, much of the inspiral is outside of the range of ground-based detectors like ET, so it is necessary to include the contributions from merger and ringdown in order to accumulate a significant SNR. The waveforms are functions of the redshifted total mass of the source, $M_z = (1+z)(M_1+M_2) \equiv (1+z)M$, and the symmetric mass ratio, $\eta \equiv M_1 M_2 / (M_1+M_2)^2$. In the present calculation we have used the coefficients from the original paper~\cite{ajith}, although updated coefficients are available~\cite{ajithUpdate}. We have verified that the SNRs computed from the two sets of coefficients are nearly identical. 

\begin{figure}[t]
\begin{center}
\includegraphics[width=0.6\textwidth, keepaspectratio=true]{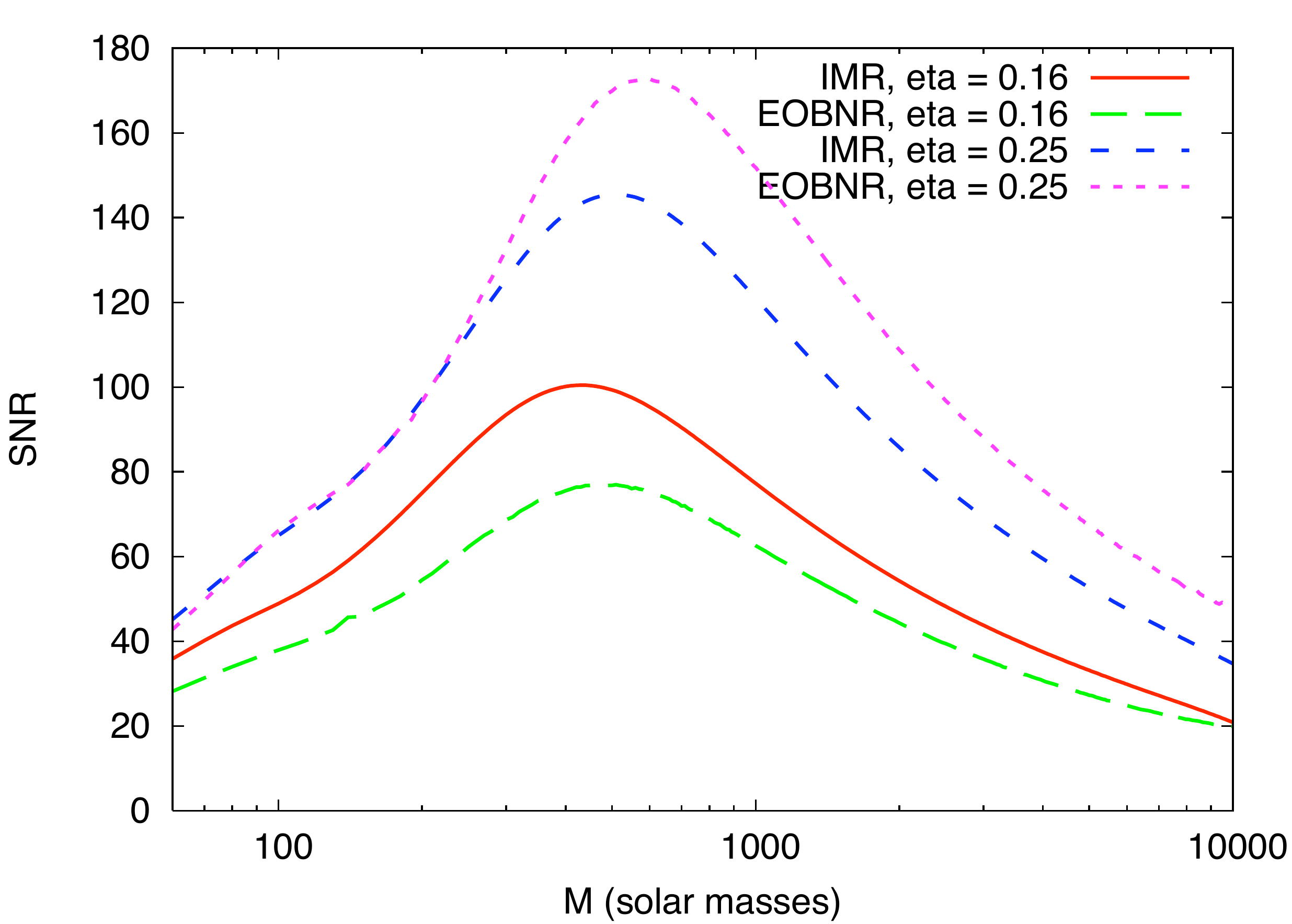}
\end{center}
\caption{Comparison of SNRs computed from IMR and EOBNR models, as a function of redshifted total mass $M_z$ and for two different choices of mass ratio $\eta=0.16$ and $\eta=0.25$. Events are at a fixed luminosity distance of $6.61$Gpc, corresponding to $z=1$, and we are computing the SNR for one 10km right-angle detector with the ET sensitivity.}
\label{f:eobvimr}
\end{figure}

As a check of our results, we also computed the SNR for events using the effective-one-body-numerical-relativity (EOBNR) waveform family introduced by Buonanno et al.~\cite{eobnr}. A comparison of the ET SNRs is shown in Fig.~\ref{f:eobvimr}. For comparable mass ratios, $\eta\sim 0.25$, the SNRs predicted by the EOBNR waveforms are somewhat higher than the IMR SNRs (by up to $\sim25\%$ when the merger and ringdown phases dominate the SNR), while for asymmetric mass ratios, $\eta \lesssim 0.16$, they tend to be somewhat lower. However, neither waveform family has been shown to be valid for $\eta < 0.16$. These comparisons give us confidence that the detection-rate estimates quoted here are reasonably trustworthy.

\subsection{Detector model}
\label{detmod}
\begin{figure}[t]
\begin{center}
\includegraphics[width=0.6\textwidth, keepaspectratio=true]{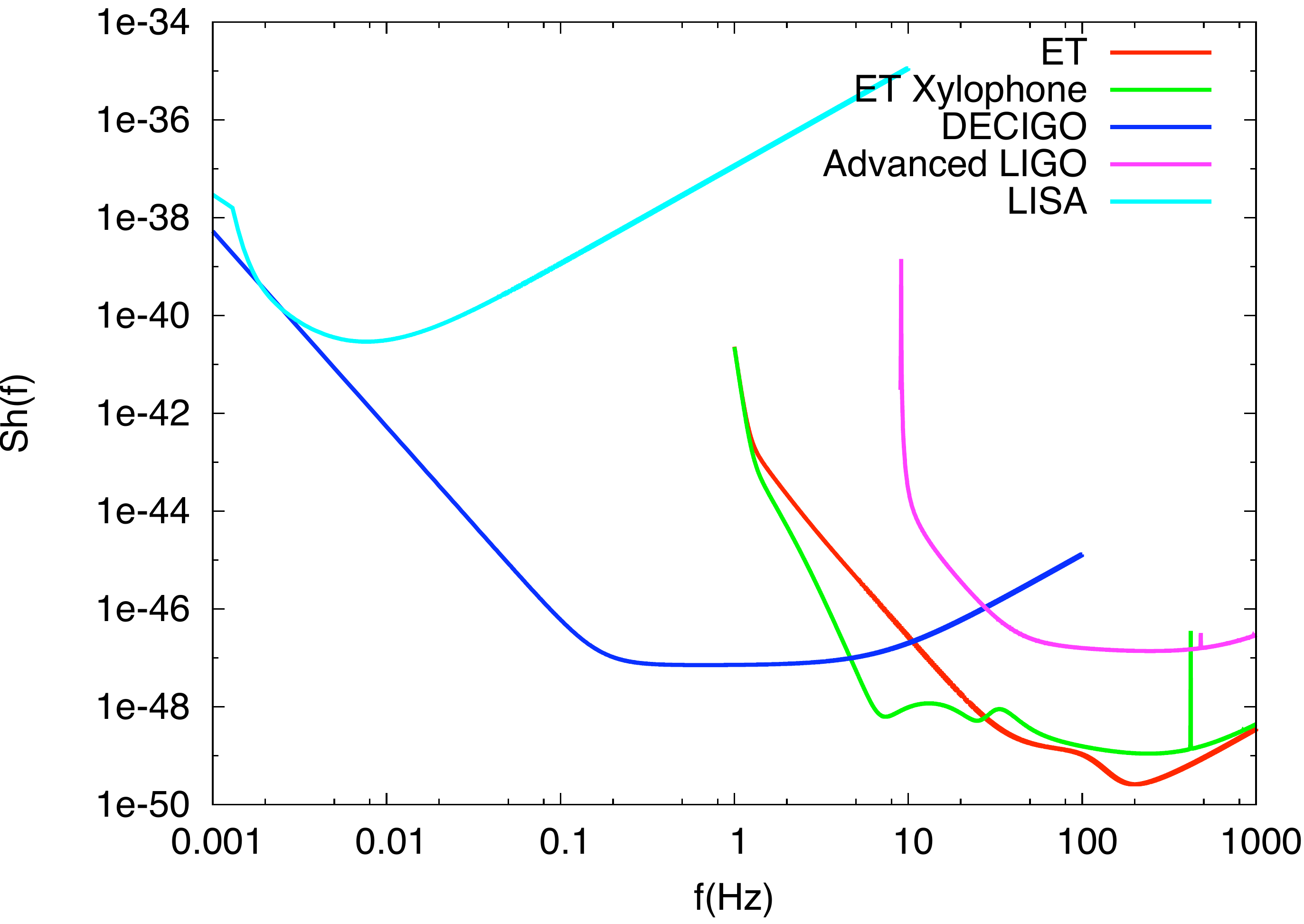}
\end{center}
\caption{Sensitivity curves for the detectors considered in this study. The LISA and DECIGO curves include a factor of $4/3$ so that they represent equivalent right-angle interferometers.}
\label{fig:noisecurve}
\end{figure}

We consider SNRs and event rates for LISA~\cite{lisa}, DECIGO~\cite{decigo}, Advanced LIGO~\cite{advligo} and ET. In addition, we present results for two different ET configurations --- a ``broad-band'' configuration as described in~\cite{freise09} which we refer to as ``ET'' and which was used to obtain the results shown in Fig.~\ref{f:eobvimr}, and a ``Xylophone'' configuration which trades off sensitivity at higher frequency for improved sensitivity near $10$Hz and was described in~\cite{etxylo}. For the SNR calculations, we compute sky- and orientation-averaged SNRs, which are obtained by computing the SNR for an optimally oriented source using Eq.~(\ref{snr}) and dividing by $2.26$. The specific detector therefore enters only through the specification of the noise spectral density. The noise spectral densities we assumed for each detector are illustrated in Fig.~\ref{fig:noisecurve}. For LISA and DECIGO the $S_h(f)$ illustrated is for an equivalent right-angle interferometer, obtained by multiplying the usual $S_h(f)$ for a 60$^{\circ}$ interferometer by $4/3$. This ensures the sky-averaging for these detectors is done in exactly the same way as the others. The DECIGO noise curve is for a single channel of one DECIGO-like interferometer. The target DECIGO design calls for four interferometers, which would enhance the SNR by a factor of 2 compared to that in a single interferometer.

The Einstein Telescope will be a 10km scale laser-interferometer. A key target in the ET design is the capability to measure polarisation at a single site. This is achievable by having two
coplanar detectors at the site, offset by 45$^{\circ}$. The currently favoured design is for a triangular facility containing three 10km 60$^{\circ}$ detectors, as this has lower facility costs, and we will
refer to this as a ``single ET''. The spectral density shown for both ET configurations in Fig.~\ref{fig:noisecurve}, and which we use in our SNR calculations, is for a single right-angle 10km detector~\cite{hild08}. The sensitivities of two right-angle 10km detectors and a ``single ET'' are factors of $\sqrt{2}$ and $3/2$ higher, respectively,  than that of one right-angle 10km detector. In Section~\ref{parest}, we will present estimates of the accuracy with which a network of ET-like detectors can measure the parameters of seed black hole mergers. We consider four ``third generation network" configurations --- i) one ET at the geographic location of Virgo, plus a second right-angle 10km detector at the location of LIGO Hanford or Perth  (Australia); ii) as configuration (i) plus a third 10km detector at the location of LIGO Livingston; iii) as configuration (i) but with the Hanford/Perth 10km detector replaced by a second ET; and iv) three ETs, one at each of the sites.

\section{Detectable events}
\label{rates}
\subsection{Sensitivity}
\begin{figure}[t]
\begin{center}
\includegraphics[width=0.6\textwidth, keepaspectratio=true]{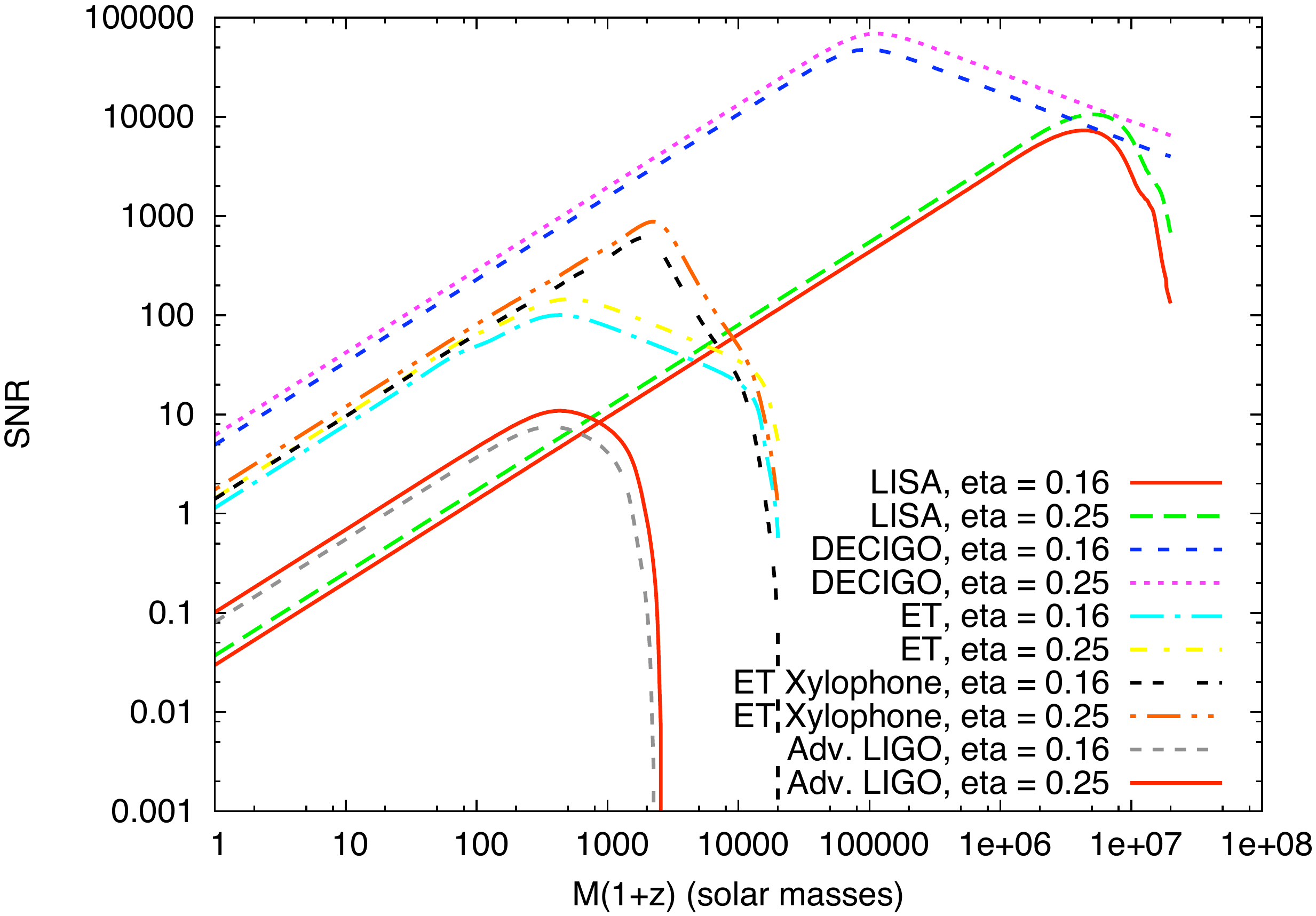}
\end{center}
\caption{Signal-to-noise ratio versus redshifted total mass, for each of the detectors and for two different mass ratios, $\eta=0.16, 0.25$. All sources are at a fixed luminosity distance of $6.61$Gpc.}
\label{fig:SNRvM}
\end{figure}

In Fig.~\ref{fig:SNRvM} we show how the SNR in each of the detectors varies as a function of the redshifted total mass for sources at fixed luminosity distance and for two different values of mass ratio, $\eta=0.16,0.25$. Advanced LIGO is unsurprisingly beaten at all masses by ET, but ET and LISA are complementary in that ET is more sensitive for $M_z<10^4M_{\odot}$, while LISA is better above that mass. This is advantageous, as these two detectors may well be operating concurrently. DECIGO beats every other detector for all masses. However, this relies on extrapolation of the DECIGO noise curve over many decades of frequency, and it is not yet clear how valid that extrapolation will be. A less ambitious second generation space-based detector like ALIA~\cite{alia} would be most sensitive for $10^3 M_{\odot} \lesssim M_z \lesssim 10^6 M_{\odot}$, but would be beaten by ET/LISA below/above that range. The xylophone configuration significantly improves ET's sensitivity for black hole mergers with $M_z\sim1000M_{\odot}$, as the instrumental noise is significantly reduced at precisely the frequency where most of the SNR is accumulated for these systems. This is very important for the seed black hole event rate. We note that the SNRs of events at redshift $z=1$ will be $\rho\lesssim10^3$ for ET, $\rho\lesssim10^4$ for LISA and $\rho\lesssim10^5$ for DECIGO.

\subsection{Event rates}
\begin{figure}[t]
\begin{center}
\includegraphics[width=0.6\textwidth, keepaspectratio=true]{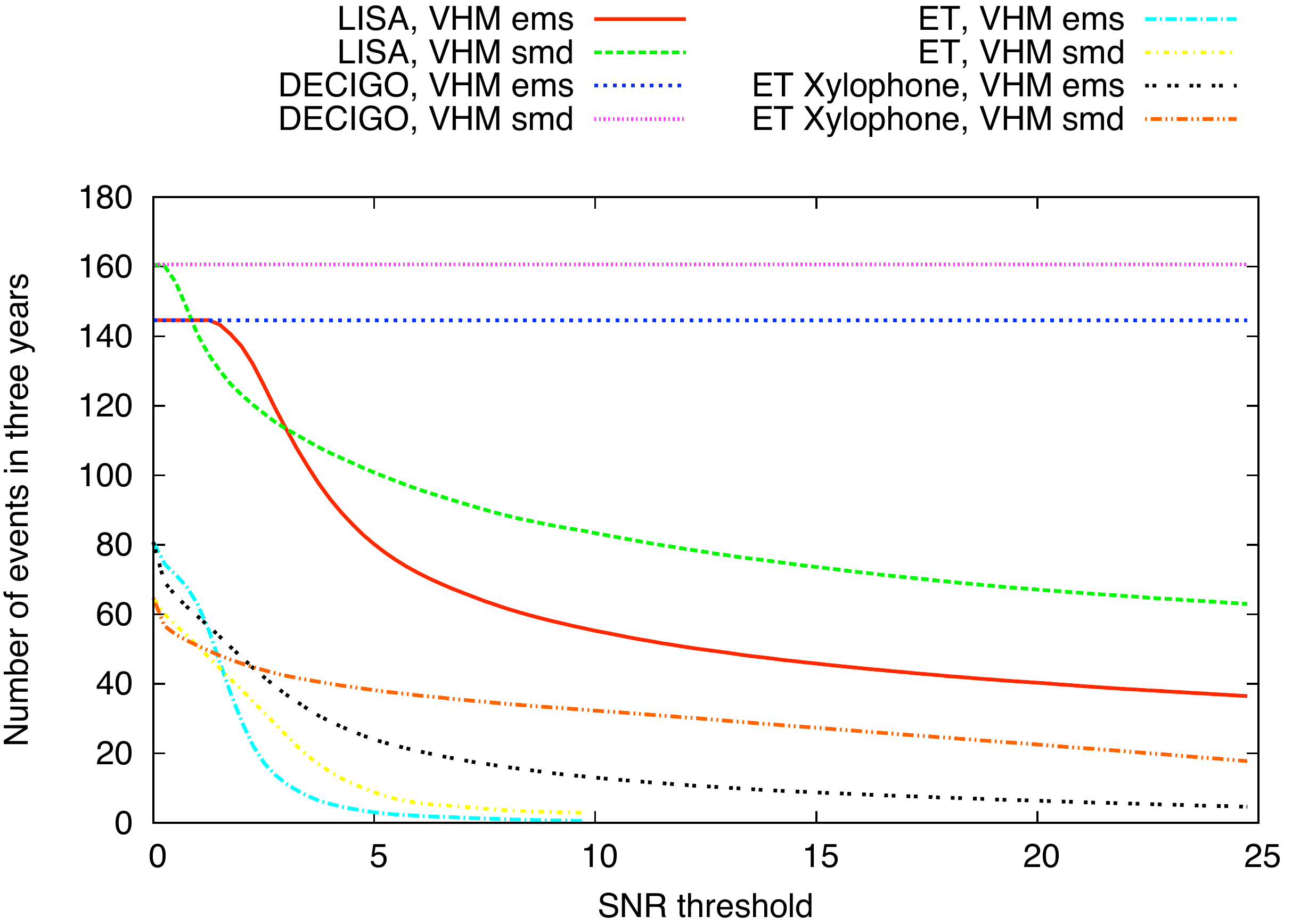}
\end{center}
\caption{Number of events detected by each detector in three years, as a function of signal-to-noise ratio threshold and for two different astrophysical models --- \emph{VHM,ems} and \emph{VHM,smd}.}
\label{fig:NvThresh}
\end{figure}

To estimate the number of events that each of the detectors would see, we used 1000 merger-tree realisations for each of the two seed black-hole initial mass-distributions described
in Section~\ref{methods} and constructed a catalogue of the merger events that occurred over three years in each case. For each event, we computed the matched filtering SNR that would be obtained for that event by each detector, using Eq.~(\ref{snr}). Specifying the SNR threshold that will be required in the detector for an event to be resolvable then determines the number of events that will be observed. The event rate is shown as a function of SNR threshold in Fig.~\ref{fig:NvThresh}. 
For the ET configurations the SNR threshold is the SNR in a single right-angle interferometer with the target sensitivity, and for LISA/DECIGO it is the SNR in a single Michelson channel. The SNR threshold that is likely to be required is $\sim 8$ for the network of detectors, which corresponds to SNRs in a single interferometer of 5.3 for a single ET, or SNRs of 4.8, 3.9, 3.8 and 3.1 for the ET network configurations (i)--(iv) described in Section~\ref{detmod}, or an SNR of 5.7 in a single Michelson channel of LISA/DECIGO (assuming that these instruments can be thought to consist of two independent data channels only). The SNR threshold of $8$ may be optimistic when other considerations are taken into account, in particular foreground confusion from other gravitational wave sources. This figure illustrates the sensitivity of the event rate to the threshold that is ultimately required.

Advanced LIGO is not included in Fig.~\ref{fig:NvThresh} because it will not detect any events in these astrophysical scenarios.  We see from Fig.~\ref{fig:NvThresh} that an ET network will detect between a few and a few tens of events, which will be mostly early mergers between light seeds; LISA will detect several tens of events, which will be mostly mergers between heavier black holes that have already undergone several merger events; and even a single DECIGO-like interferometer will provide a complete survey of these events, which is why the DECIGO curves are flat for all values of the SNR threshold shown in the plot. We note that the ET xylophone configuration would detect significantly more seed black hole events at relatively high SNR than the broadband ET configuration. This should be an important consideration when final design decisions are made. We emphasise that these results should not be considered to be robust predictions for the event rates, due to the uncertainties in the astrophysical models, but are designed to illustrate the potential science that these future detectors could do.

\subsection{Mass/redshift distribution of detected events}
\begin{figure}[t]
\begin{center}
\begin{tabular}{cc}
\includegraphics[width=0.51\textwidth, keepaspectratio=true]{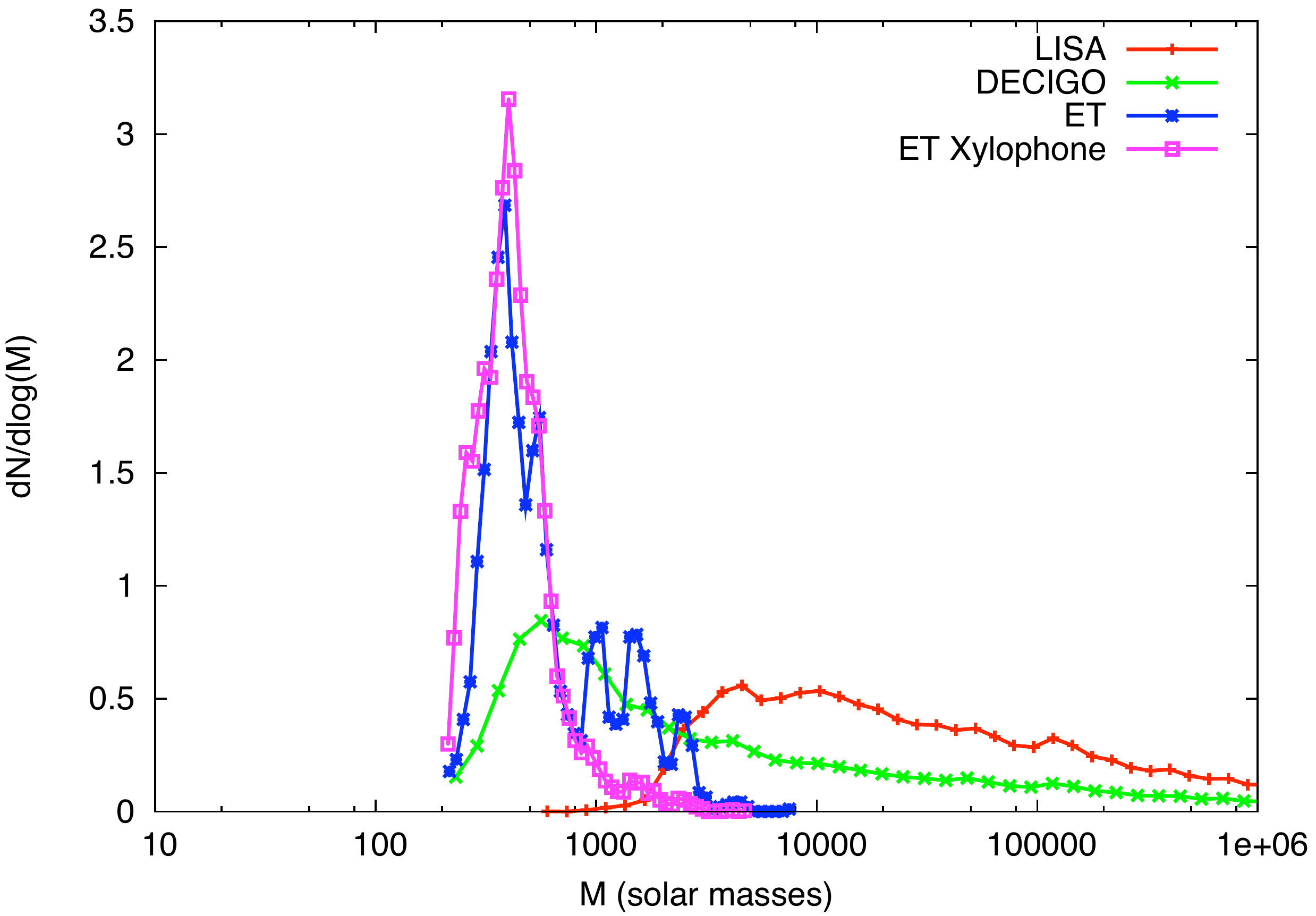}&\includegraphics[width=0.5\textwidth, keepaspectratio=true]{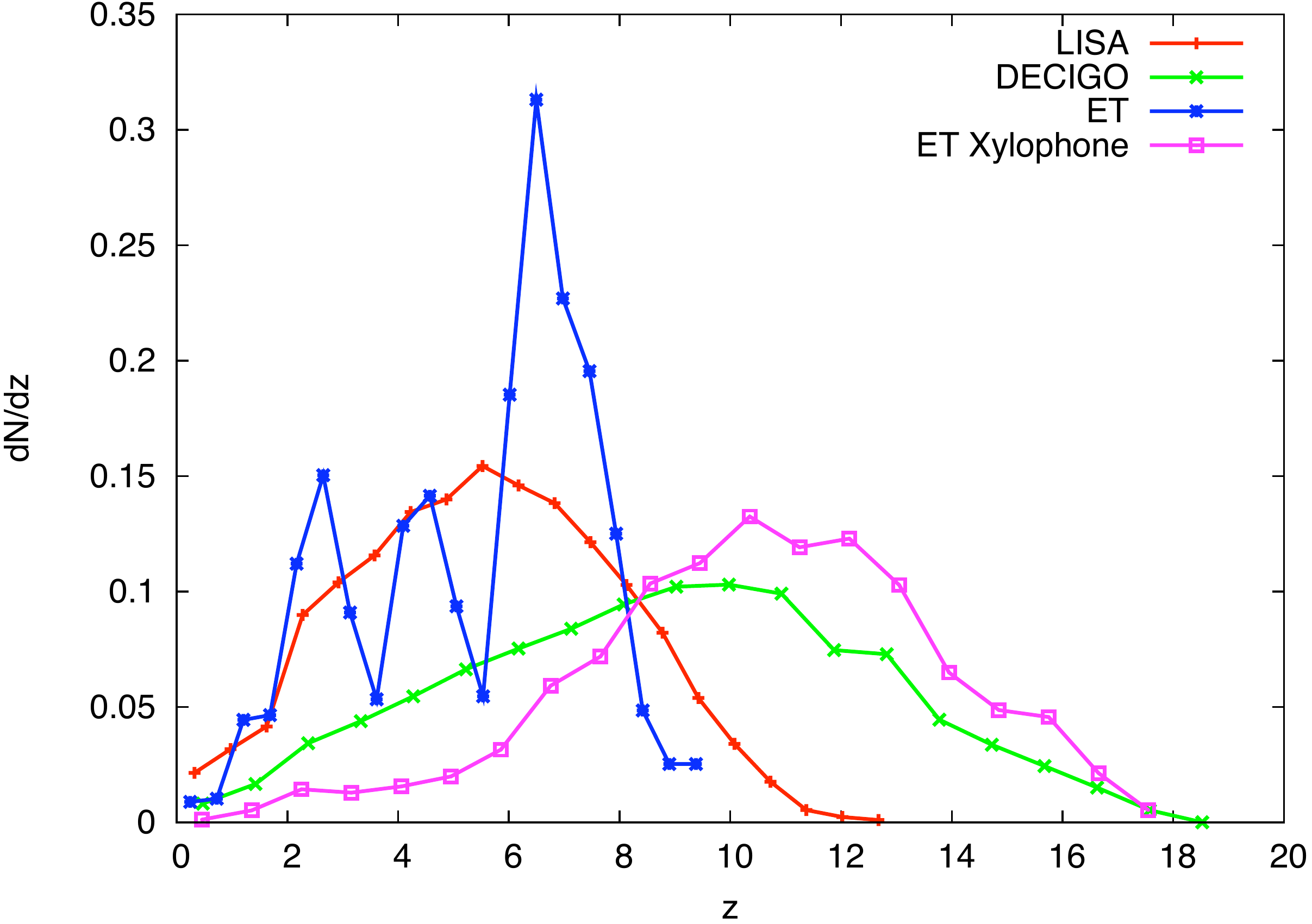}\\
\includegraphics[width=0.51\textwidth, keepaspectratio=true]{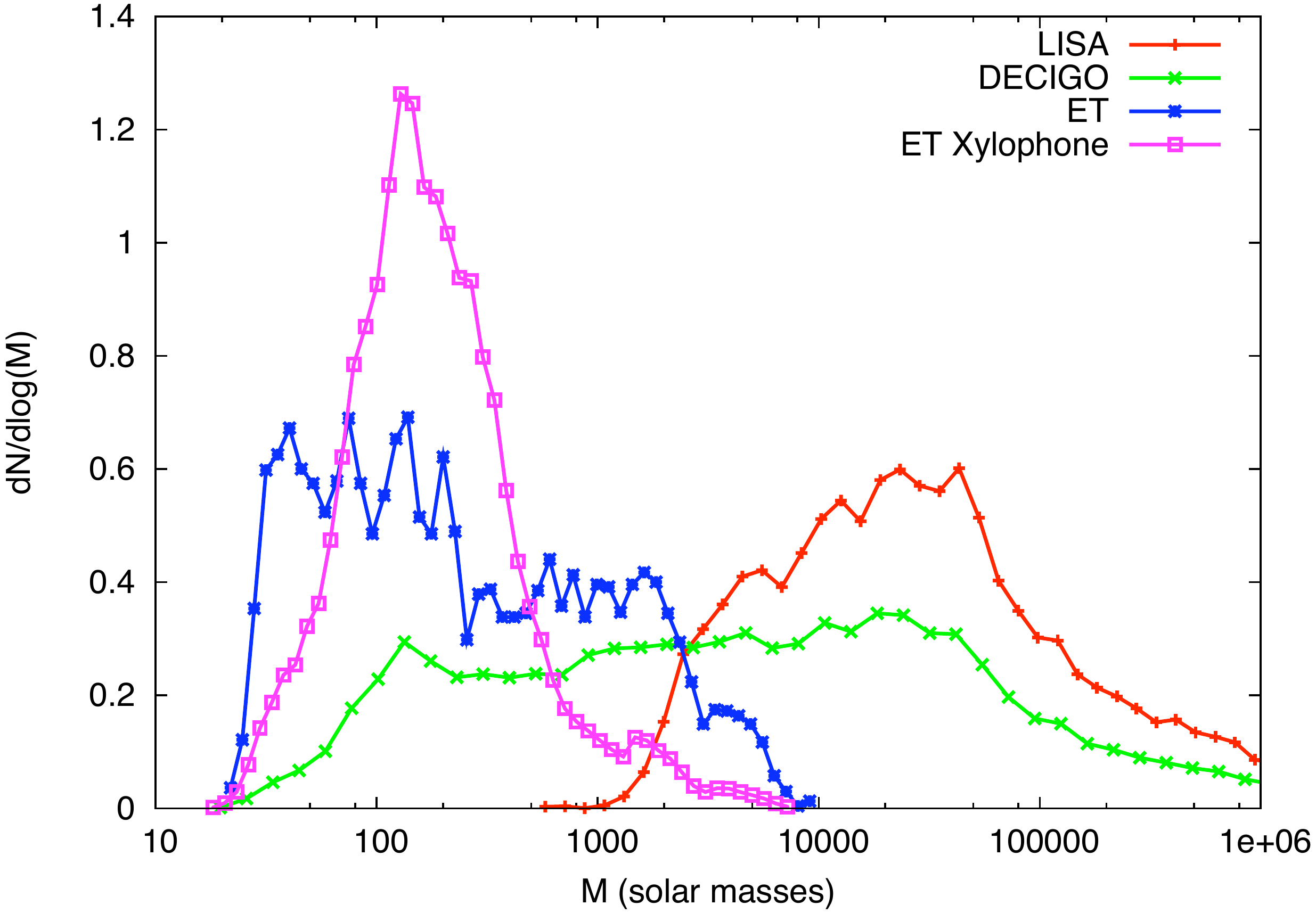}&\includegraphics[width=0.5\textwidth, keepaspectratio=true]{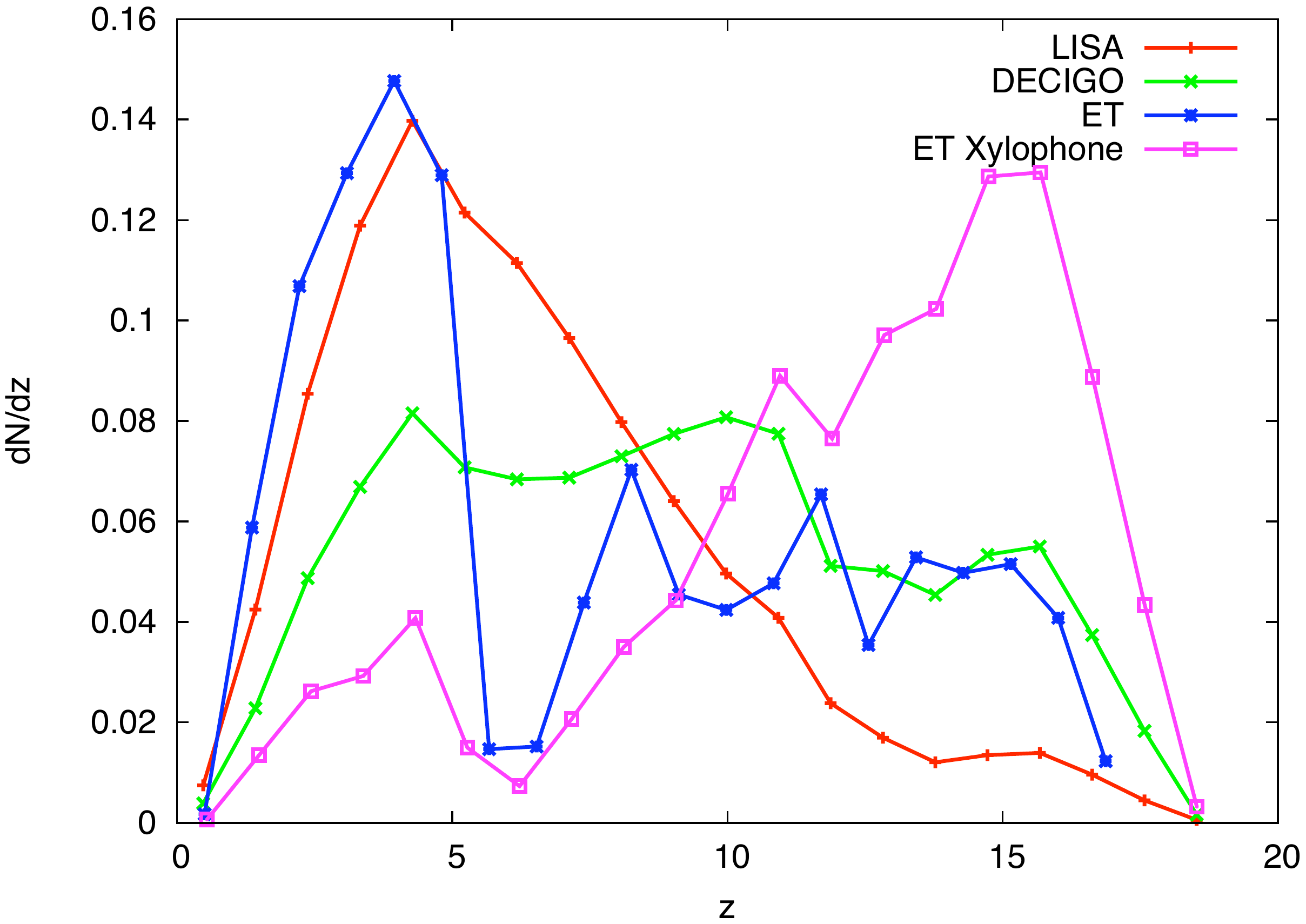}
\end{tabular}
\end{center}
\caption{Intrinsic mass (left panels) and redshift (right panels) distributions of events for each detector, for the \emph{VHM,ems} scenario (upper panels) and the \emph{VHM,smd} scenario (lower panels). These were computed using all events with SNR greater than $5$ in a single 10km right-angle interferometer for the Einstein Telescope and SNR greater than $10$ in a Michelson channel for LISA. The larger SNR threshold was used for LISA since we will not have the benefit of multiple detectors in that case. The choice of SNR threshold for DECIGO is not important since all events are detected with SNR greater than $25$.}
\label{fig:dNdM}
\end{figure}

The event rate alone does not provide a proper comparison between the different detectors, since it suppresses information about the characteristics of the events. In Fig.~\ref{fig:dNdM} we show the mass and redshift distributions of events detected by each instrument for the \emph{VHM,ems} and \emph{VHM,smd} scenarios. Each curve is normalised so that the integral under the curve is equal to 1, and the mass distribution is for the intrinsic mass, $M$, not the redshifted mass. These plots reflect the complementarity of the detectors noted earlier. LISA and ET are almost completely complementary in mass --- ET/LISA detects events only below/above $\sim 2000M_{\odot}$ --- but probe a comparable range in redshift. The distribution of DECIGO events follows the true astrophysical distribution of mergers --- most mergers are between $\sim500M_{\odot}$ black holes in the \emph{VHM,ems} scenario, but the mass distribution is quite flat in the \emph{VHM,smd} case, in the range $\sim10^2-10^5M_{\odot}$. In both cases the redshift distribution of events is quite flat and extends up to $z\approx15$. If DECIGO was replaced by ALIA, then it would not be able to detect the tail of $M\lesssim 100M_{\odot}$ black holes found in the \emph{VHM,smd} model, leaving only ET able to probe those systems. ET events are predominantly of low mass $\sim10$--$1000M_{\odot}$ in both configurations and would have redshift $z<10$ if it was operated in the broad band mode; if ET was operated in the xylophone configuration, it would be able to detect mergers out to $z\sim15$ and indeed would be dominated by events with $10 < z < 15$. LISA events are predominantly of high mass $\sim10^3$--$10^6 M_{\odot}$ and redshift $z<10$.
The coalescences detected by LISA are between black holes that have already undergone several merger events, while ET probes the earliest mergers between seed black holes. Thus, LISA and ET together will probe the full merger history of galaxies in these scenarios. DECIGO or ALIA by themselves would have the same capability, which is an important scientific motivation for building these ambitious instruments.

\section{Accuracy of parameter-estimation}
\label{parest}
It is known that LISA (and DECIGO) will provide highly accurate measurements of the parameters of the long-lived mergers of black holes with $M \gtrsim 10^5M_{\odot}$~\cite{lisape}. However, for the first generation of mergers with $M \lesssim 10^3M_{\odot}$, the source spends little time in the sensitive frequency band of any of the detectors.  The intrinsic parameters (masses etc.) of the source affect the waveform phasing and hence should be determined accurately from the observed waveform. However, the waveform  at the detector also depends on six extrinsic parameters --- two sky-position angles,  the orbital phase at a fiducial time $t_0$, the wave polarization angle $\psi$, the source inclination angle with respect to the line of sight $\iota$, and the luminosity distance to the source $D_L$. These affect only the amplitude of the response in the detector. Any number of colocated and coplanar detectors provides only four amplitude measurements for a short-lived event --- two quadratures in two detectors at 45 degrees to each other. Thus, to measure all the source parameters, we need at least one other non-colocated detector operating concurrently. The full DECIGO mission aims to have four interferometers, two of which would be colocated in a ``star-of-David'' configuration, to allow stochastic background measurements. If this configuration is realised, the DECIGO network will be able to make high-precision parameter estimates for seed black hole mergers. On the ground, parameter estimates could be obtained from a networks of ETs, as described in Section~\ref{detmod}, with up to three detectors sited at the geographical locations of Virgo, LIGO Hanford and LIGO Livingston (VHL) or Virgo, Perth and LIGO Livingston (VPL). 

\begin{figure}[t]
\begin{center}
\includegraphics[width=0.6\textwidth, keepaspectratio=true]{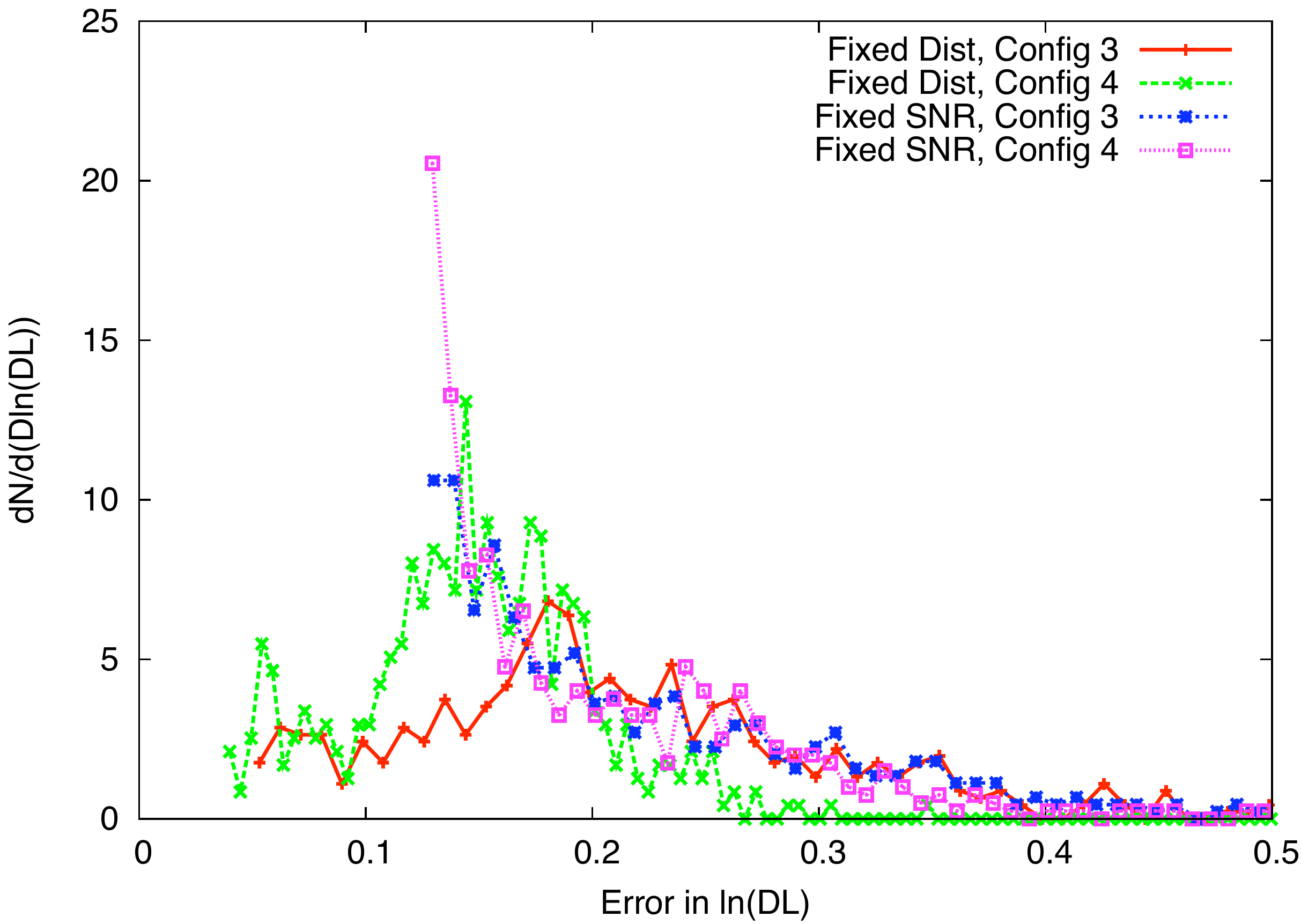}
\end{center}
\caption{Distribution of fractional distance errors for sources with $M_z=100\msun$, $\eta=0.15$, estimated from a Monte Carlo simulation over the extrinsic parameters. Curves are shown for network configurations (iii) and (iv) and for two separate cases --- sources with fixed SNR $\rho=8$, and sources at a fixed distance of $D_L=10$Gpc.}
\label{fig:DistErrs}
\end{figure}

In order to assess the accuracy of parameter estimates that would be obtained by a ground-based network, we have performed a Monte Carlo over possible choices of the extrinsic parameters, for fixed intrinsic parameters. We carried out two different simulations --- one in which the SNR of the source was fixed at $\rho=8$ and one in which the luminosity distance to the source was fixed. For these simulations, we used the standard broad-band ET noise curve rather than that of the xylophone configuration. In Tables~\ref{tabparest}--\ref{tabparestFixedDist} we summarise these results by listing the error at the upper 68$^{\rm th}$ percentile of the distribution derived from the Monte Carlo simulation. We consider the upper percentile only since we are only concerned about errors that are particularly large. In Fig.~\ref{fig:DistErrs} we also show the full distribution of the errors in the luminosity distance computed from the simulations. As expected, we find that the redshifted mass $M_z$ and the symmetric mass ratio $\eta$ are determined very well, to an accuracy better than $1\%$ even for sources near the threshold $\rho = 8$. The variation in these errors in the fixed-distance case comes entirely from differences in SNR as the extrinsic parameters vary. We should note, however, that the physical waveforms are likely to be different from the simple model waveforms we have used for parameter estimation here, and will include additional features such as spins
and associated precessions.  Ignoring such features could lead to
significant systematic errors, while including them in the waveform
family used for parameter estimation would likely create additional
correlations between parameters leading to a degraded
parameter-estimation accuracy (see, e.g., \cite{sluys09}). The distance is not so well determined, and the error distribution is highly non-Gaussian. For some sources, we can achieve the theoretical best accuracy of $1/\rho$, indicated by the hard cut-off at $12.5\%$ in the fixed SNR distribution. While there is a tail out to large distance errors of $100\%$ or more, for the majority of cases the errors are smaller than $50\%$. Our results also allow us to make statements about the network configuration. We see that only one additional right-angle interferometer is required to achieve $\lesssim 40\%$ distance precision. Adding a third detector or upgrading the detectors to ETs only modestly improves these errors to $\lesssim 25\%$ for events close to the SNR threshold, although the SNR, and hence parameter precision, for a source at a given distance will be significantly improved.  Finally, we note that it does not seem to matter whether the second detector is at Hanford or in Australia.

To identify an event as a merger between seed black holes, we need to be able to say that the mass is low, $M\lesssim 10^3M_{\odot}$, and the redshift is high, $z \gtrsim 3$. Intermediate-mass black hole binaries could also form at high redshift through runaway stellar collisions in globular clusters (for more discussion see~\cite{apjpaper}). While it is not clear at present if it will be possible to distinguish between events involving black holes formed via these two channels, the distinction between the mechanisms becomes increasingly arbitrary at higher redshift. What is important for the light-seed versus heavy-seed debate is to know that $\sim 100M_{\odot}$ black holes existed at high redshift. The error in $M$ is dominated by the error in $z$, since we determine the redshifted mass $M_z = (1+z)M$ very precisely. We expect to calculate $z$ from the recovered $D_L$ and the concordance cosmology at the time of the observations, so the error in $z$ and hence $M$ will be comparable to that in $D_L$, i.e., $\lesssim 40\%$. Thus we should be able to confidently say that the events with $M\sim 100M_{\odot}$ and $z\sim5$ which dominate the ET detection rate are indeed seed merger events. One caveat for these conclusions is that the distance errors computed here were based on the Fisher Matrix, which is known to overestimate the precision of measurements in the low SNR limit~\cite{vallis08}. They should therefore be regarded as optimistic. A proper calculation would require Monte Carlo simulations to recover the full posterior for a signal injected into a noisy data stream, which is beyond the scope of the present work. However, the results here provide a guide to what might be achievable in practice.

\begin{table*}
{\footnotesize\begin{tabular}{|c|c||c|c|c|c|c|c|c|c|c|c|}
\hline&&&&\multicolumn{4}{c|}{$\Delta{\ln D_L}$ (VHL)}&\multicolumn{4}{c|}{$\Delta{\ln D_L}$ (VPL)} \\\cline{5-12} $M_z$&$\eta$&$\Delta{\ln M_z}$&$\Delta{\ln \eta}$&(i)&(ii)&(iii)&(iv)&(i)&(ii)&(iii)&(iv)\\\hline
$100\msun$&$0.15$&$0.1\%$&$0.05\%$&$ 37\%$&$ 27\%$&$ 25\%$&$ 23\%$&$36\%$&$27\%$&$25\%$&$23\%$\\\hline
$100\msun$&$0.25$&$0.2\%$&$0.06\%$&$ 37\%$&$ 28\%$&$ 24\%$&$ 22\%$&$36\%$&$26\%$&$24\%$&$22\%$\\\hline
$500\msun$&$0.15$&$0.9\%$&$0.4\%$&$ 41\%$&$ 31\%$&$ 29\%$&$ 25\%$&$41\%$&$30\%$&$28\%$&$26\%$\\\hline
$500\msun$&$0.25$&$0.1\%$&$0.1\%$&$ 37\%$&$ 32\%$&$ 28\%$&$ 24\%$&$35\%$&$30\%$&$28\%$&$26\%$\\\hline
$1000\msun$&$0.15$&$2\%$&$1\%$&$ 53\%$&$ 33\%$&$ 31\%$&$ 26\%$&$43\%$&$33\%$&$32\%$&$27\%$\\\hline
$1000\msun$&$0.25$&$0.3\%$&$0.1\%$&$ 42\%$&$ 31\%$&$ 31\%$&$ 27\%$&$34\%$&$30\%$&$30\%$&$26\%$\\\hline
\end{tabular}}
\caption{Parameter-estimation accuracy for various ET-network configurations for sources at a fixed signal-to-noise ratio of $8$. The values reported for $M_z$ and $\eta$ are statistical averages, while those for $D_L$ are the values at the $68^{\rm th}$ percentile of the error distribution.
\label{tabparest}
}
\end{table*}


\begin{table*}
{\footnotesize \begin{tabular}{|c|c||c|c|c|c|c|c|c|c|c|c|c|c|}
\hline&&\multicolumn{4}{c|}{$\Delta{\ln M_z}$}&\multicolumn{4}{c|}{$\Delta{\ln \eta}$}&\multicolumn{4}{c|}{$\Delta{\ln D_L}$} \\\cline{3-14} $M_z/\msun$&$\eta$&(i)&(ii)&(iii)&(iv)&(i)&(ii)&(iii)&(iv)&(i)&(ii)&(iii)&(iv)\\\hline
$100$&$0.15$&$0.12\%$&$0.11\%$&$0.11\%$&$0.09\%$&$0.008\%$&$0.007\%$&$0.007\%$&$0.006\%$&$37\%$&$26\%$&$24\%$&$17\%$\\\hline
$100$&$0.25$&$0.15\%$&$0.13\%$&$0.13\%$&$0.1\%$&$0.01\%$&$0.01\%$&$0.01\%$&$0.009\%$&$26\%$&$19\%$&$17\%$&$12\%$\\\hline
$500$&$0.15$&$0.5\%$&$0.46\%$&$0.45\%$&$0.4\%$&$0.04\%$&$0.03\%$&$0.03\%$&$0.03\%$&$17\%$&$13\%$&$13\%$&$9\%$\\\hline
$500$&$0.25$&$0.05\%$&$0.04\%$&$0.04\%$&$0.03\%$&$0.01\%$&$0.009\%$&$0.009\%$&$ 0.007\%$&$12\%$&$8\%$&$8\%$&$6\%$\\\hline
$1000$&$0.15$&$2\%$&$1.6\%$&$1.6\%$&$1.3\%$&$0.14\%$&$0.1\%$&$0.1\%$&$0.09\%$&$30\%$&$19\%$&$18\%$&$12\%$\\\hline
$1000$&$0.25$&$0.17\%$&$0.14\%$&$0.14\%$&$0.11\%$&$0.02\%$&$0.01\%$&$0.01\%$&$0.01\%$&$31\%$&$10\%$&$11\%$&$7\%$\\\hline
\end{tabular}}
\caption{As Table~\ref{tabparest}, but now for sources at a fixed distance of $10$Gpc. In this case, since the SNR varies with choices of extrinsic parameters, we quote errors for $M_z$, $\eta$ and $D_L$ as the values at the $68^{\rm th}$ percentile of the error distribution. We quote only errors for the VPL configuration, as those for the VHL configuration are very similar.}
\label{tabparestFixedDist}
\end{table*}

\section{Discussion} 
\label{discuss}
In this paper we have explored the capability of future gravitational-wave detectors to probe the hierarchical assembly of structure in the Universe starting from light seeds of massive black holes,
born during supernovae of population III stars. In the astrophysical scenarios that we considered, LISA would detect several tens of events, but cannot probe black hole masses below $\sim 1000M_{\odot}$. While these observations will have some power to distinguish between light-seed and heavy-seed scenarios (see, e.g.,~\cite{plowman}), LISA cannot probe the first generation of mergers. The Einstein Telescope will complement LISA by detecting a few to a few tens of black hole mergers per year in the $10$--$10^3M_{\odot}$ range, detecting the first epoch of mergers and thus directly probing the mass-distribution of the seeds.  DECIGO would provide a complete survey of MBHB mergers occurring over its lifetime. The launch date for this instrument is rather uncertain, but if it achieves the target of ~2024 it may operate at the same time as ET, allowing simultaneous measurements. More likely, DECIGO will be later than this, in which case the detections it makes will confirm earlier results from LISA and ET. These future observations will provide detailed information on the assembly of structure and stringent constraints on viable models of galaxy formation and growth. The detection of at least one light seed will itself be of huge significance in constraining the heavy-seed model. Of the two alternative designs for ET that we have considered, it is quite clear that the xylophone configuration is to be preferred for the detection of these sources. It not only enhances the event rate, but it allows ET to detect mergers at higher redshift, which is important for having confidence that the events are seed mergers. This should be born in mind when the configuration for ET is finalised.

We have also discussed parameter estimation for these merger events. Measuring the parameters of the short-lived events with $M \lesssim 10^3M_{\odot}$ will require a network of concurrently operating detectors. We have shown that a ground-based network of ET detectors should be able to measure the redshifted mass and the luminosity distance of the majority of light-seed merger events to accuracies of $\lesssim 1\%$ and $\sim 40\%$ respectively. This should be sufficient for us to say with confidence that we are observing mergers involving light seeds of massive black holes at high redshift.

Our results indicate that Advanced LIGO will not detect any events from these sources. This is in contrast to a prediction made in~\cite{wyithe}. However, that previous calculation was based on a semi-analytic prescription for the black hole merger history which overpredicts the number of events involving low-mass black holes at low redshift as compared to our merger-tree calculations. In the model of~\cite{wyithe}, ET would detect several hundred events rather than several, which our calculations suggest is very optimistic.

The current work should not be taken as a robust calculation of the properties of the events that will be observed by future gravitational-wave detectors, but as an indication of the potential science that these detectors will do. More work is needed to properly quantify exactly what gravitational-wave observations will tell us about the growth of structure. Nonetheless, we expect that the 
calculations described here will be of use to researchers interested in quantifying such questions in the future.

\ack
We would like to thank M.~Volonteri for useful discussions and for providing the Monte-Carlo realisations of the halo and MBH merger hierarchy.  We also thank A.~Buonanno, P.~Ajith, Y.~Pan and E.~Ochsner for very helpful discussions of the IMR and EOBNR waveforms.  We used code from the LSC Algorithm library for the EOBNR waveform family.  JG's work is supported by the Royal Society. IM is partially supported by NASA ATP Grant NNX07AH22G to Northwestern University. AS acknowledges support from National Science Foundation Grants No.~PHY 06-53462 and No.~PHY 05-55615, and NASA Grant No.~NNG05GF71G, awarded to The Pennsylvania State University.  AV is partially supported by the UK Science and Technology Facilities Council. 

{}
\end{document}